\documentclass[prb,aps,twocolumn,groupedaddress,floats,showpacs]{revtex4-1}
\usepackage{graphicx}
\usepackage{dcolumn}
\usepackage{bm}
\def\gapx{\lower 2pt \hbox{$\buildrel>\over{\scriptstyle{\sim}}$\ }}
\def\lapx{\lower 2pt \hbox{$\buildrel<\over{\scriptstyle{\sim}}$\ }}

\begin{document}

\title{Phase diagram of soft-core bosons in two dimensions}

\author {S. Saccani$^1$, S. Moroni$^{1}$ and M. Boninsegni$^2$}

\affiliation {$^1$SISSA Scuola Internazionale Superiore di Studi Avanzati and
        DEMOCRITOS National Simulation Center,
          Istituto Officina dei Materiali del CNR
         Via Bonomea 265, I-34136, Trieste, Italy
}
\affiliation {$^2$Department of Physics, University of Alberta, Edmonton, Alberta, Canada T6G 2G7}
\date{\today}
\begin{abstract}
The low temperature phase diagram of Bose soft disks in two dimensions is studied by numerical simulations. It is shown that a supersolid cluster phase exists,  within a range of the model parameters, analogous to that recently observed for a system of aligned dipoles interacting via a softened potential at short distance. These findings indicate  that a long-range tail of the interaction is unneeded to obtain such a phase, and that the soft-core repulsive interaction is the minimal model for supersolidity.
\end{abstract}


\pacs{67.80.K-, 67.85.Hj, 67.85.Jk, 67.85.-d, 02.70.Ss}

\maketitle

{\em Introduction}.
The supersolid phase of matter, displaying simultaneously crystalline order and dissipation-less flow, is a subject of long standing interest in condensed matter and quantum many-body physics. In recent years, the attention of theorists and experimenters alike has focused on solid $^4$He, following the observation of non-classical rotational inertia by Kim and Chan.\cite{KC}  At the present time, agreement is still  lacking, as to whether experimental findings indeed mark the first observation of supersolid behaviour.\cite{nikolay07} The most reliable theoretical studies, based on first-principle numerical simulations, show that superfluidity if it occurs at all in solid helium, is not underlain by the mechanism originally envisioned in the seminal works by Andreev, Lifshitz and Chester, i.e., through Bose Condensation of a dilute gas of vacancies or interstitials,\cite{andreev,chester} but involves instead extended defects, such as dislocations.\cite{pollet08} In particular, a dilute gas of point defects in solid helium has been predicted to be thermodynamically unstable.\cite{boninsegni07}

Regardless of how the current controversy over the interpretation of the present $^4$He phenomenology is eventually resolved, it seems fair to state that solid helium does not afford a direct, simple, and clear observation of the supersolid phenomenon. Still, among all simple atomic or molecular condensed matter systems, helium should be the best candidate by far, due to the favourable combination of large quantum delocalization of its constituent (Bose) particles, and weakness of the interatomic potential. But what exactly, in the physics of this simple crystal, contributes  to suppress (if not eliminate entirely) its superfluid response ?

The thermodynamics of solid $^4$He, as it emerges from first-principle quantum simulations, is largely 
determined by the strong repulsive core of the pair-wise interatomic potential at short distance.
For example, a very simple model of Bose hard spheres reproduces surprisingly accurately the phase diagram of condensed helium. Such a repulsive core is a ubiquitous feature of ordinary  interactions between atoms or molecules, arising from the Pauli exclusion principle, acting between electronic clouds of different atoms. Computer simulation studies of classical crystals, making use of the Lennard-Jones potential, have yielded evidence of the same  instability of a gas of point defects (vacancies) observed in the quantum system.\cite{ma} This suggests that the origin of such instability may lie in the strong interaction (specifically the hard core repulsion) among particles, which quantum delocalization cannot overcome. 
One is then led to pose the theoretical question of which type of inter-particle interaction (or, class thereof) might underlie supersolid behaviour. In particular, can an interaction featuring a ``softer" core, saturating at short distance to a value of the order of  the characteristic zero-point kinetic energy of the particles, result in the appearance of a supersolid phase ?
\\
This question might have seemed little more than ``academic" until not so long ago, for how would one go about creating artificially such an interaction, which does not occur in any known naturally occurring quantum many-body system ?
However, impressive advances in cold atom physics  appear to allow one to do just that, namely  to ``fashion" artificial inter-particle potentials, not arising in any known condensed matter system. It makes therefore sense to search theoretically for supersolid, or other exotic phase of matters, based on more general types of interactions among elementary constituents than the ones considered so far, with the realistic expectation that such interactions might be realizable in the laboratory.

In a recent paper, \cite{cinti} a two-dimensional system of bosons was studied, interacting via a purely repulsive pair-wise potential, decaying as  1/$r^3$ at long distance but saturating to a finite value as particles approach one another. This particular form of potential might be feasible in cold atomic systems, through a mechanism known as Rydberg blockade.\cite{lukin}
Such a system  displays at low temperature a crystalline phase in which unit cells feature more than one particle. In turn, such a crystal turns superfluid at sufficiently low temperature, phase coherence being established by quantum-mechanical tunnelling of particles across adjacent lattice sites.

Multiple occupancy crystals (or cluster crystals) are a well known subject in classical physics, as a model for polymers interaction.\cite{Mladek1} In particular, a criterion is known for the quantitative prediction of clustering, based on the form of the molecular interatomic potential.\cite{Likos1}  It is reasonable to expect that the basic physics should remain relevant for a quantum-mechanical system as well. However, it is not clear what role, if any, the long-range repulsive tail of the interaction plays, in the occurrence of superfluidity of the cluster crystal. 

In Ref. \onlinecite{cinti}, supersolid cluster crystal phases were observed in numerical simulations with long-range tails other than 1/$r^3$ (e.g., 1/$r^6$), and indeed it is simple to convince oneself that a long-range tail is not required, in order for a cluster crystal phase to exist. For example, the classical ground state of a system of particles interacting via the following, {\it soft core} potential
\begin{eqnarray}
v(r) = \left\{
\begin{array}{l l}
  V & \quad \mbox{if $r \leq a$}\\
  0 & \quad \mbox{if $r > a$}\\
\end{array} \right. 
\label{Table}
\end{eqnarray}
will be a cluster crystal at densities for which the mean inter-particle distance $d$ is less than the soft-core diameter $a$. The average number of particles $K$ per cluster can also be easily established  to be equal to $\gamma a^2/d^2$, where $\gamma$ is a number (slightly greater than one) that depends on dimensionality. In other words, $K$ is independent of $V$.

In order to investigate in greater detail the importance played by the tail of the inter-particle interaction in the stabilization of a supersolid cluster solid, as well as to contribute to the search for the ``minimal model" of  supersolidity, we have investigated in this work the low temperature properties of a two-dimensional system of Bose {\it soft disks}, i.e., particles interacting via the simple potential given by Eq. (\ref{Table}). In spite of its simplicity, to our knowledge (and surprise) this has not been the subject of any prior theoretical study. Our calculations are numerical, based on the Continuous-space Worm Algorithm. 
\\
Our main result is that the same phase(s) observed in Ref. \onlinecite{cinti} are present in the system considered here, which is therefore arguably the simplest model system underlying a supersolid phase. Because of the substantial irrelevance of the long-range form of the repulsive tail, we may conclude that the supersolid cluster crystal phase should be observable in a relatively broad class of interactions.

{\em Model and Methodology}.
We consider a system of Bose particles of spin zero in two dimensions. 
The Hamiltonian of the system in reduced units is
\begin{eqnarray}\label{mod}
\mathcal{H}=-\frac{1}{2}\sum_{i=1}^N \mathbf{\bigtriangledown}_{i}^2+ D \sum_{i>j} \Theta(|r_{ij}-1|),
\end{eqnarray}
where $r_{ij}$ is the distance between particles $i$ and $j$, all distances are expressed in units of the soft-core diameter $a$, while all energies are expressed in units of $\epsilon_\circ = {\hbar^2}/{ma^2}$. The parameter $D \equiv V/\epsilon_\circ$ can also be expressed as $(a/\xi)^2$, where $\xi$ is the quantum-mechanical penetration length of a potential barrier of height $V$.  In the limit $\xi\to 0$ the model (\ref{mod}) reduces to the hard-sphere gas.

Quantum Monte Carlo simulations of the system described by (\ref{mod}) have been performed by means of the Continuous-space Worm Algorithm, in the grand canonical ensemble (i.e., at fixed temperature $T$, area $A$ and chemical potential $\mu$). Because  this methodology is by now well-established, and is thoroughly described elsewhere, we omit here technical details, and refer the interested reader to the original references.\cite{worm,worm2} The system is enclosed in a cell with periodic boundary conditions. We denote by $N$ the average number of particle and express the density $\rho$ in terms of the dimensionless parameter $r_s=1/\sqrt{\rho a^2}$.

{\em Results}. We present here results obtained varying  $\mu$ and $D$, in the $T\to 0$ limit; that is, in all cases shown explicitly, the value of the temperature is sufficiently low that estimates can be regarded as essentially ground state ones (typically $T\sim \epsilon_\circ$, for most quantities).
\begin{figure}

 \includegraphics[width=\columnwidth]{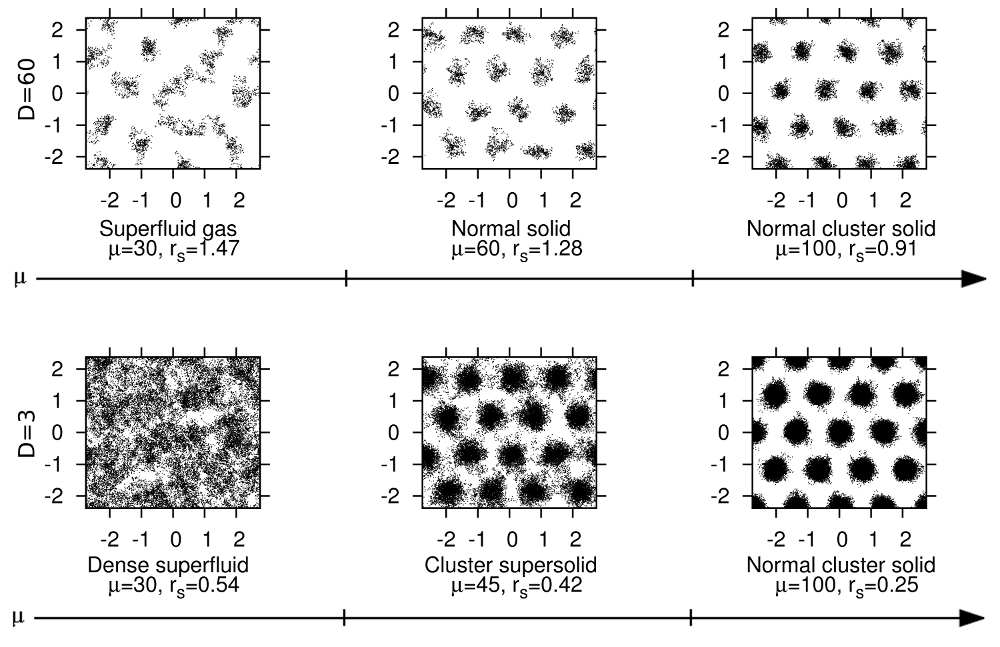}

\caption{\label{Fig_1} Qualitative low temperature phase diagram for high and low $D$ as a function of $\mu$. The panels show typical spatial configuration of the world lines resulting from simulations at low $T$, referring to the various phases. Results shown in the upper part of the figure correspond to simulations with $D=60$, whereas the lower part to $D$=3.}
\end{figure}
 
Figure \ref{Fig_1} summarizes qualitatively the ground state phase diagram in the $D-\mu$ plane, with the aid of  instantaneous (in the Monte Carlo sense) many-particle configurations (i.e., world lines). For $D >> 1$ ($D=60$ in the figure), the physics of the system is essentially that of the hard-sphere fluid. At low density, the system is a superfluid gas, which undergoes solidification into a triangular crystal on increasing $\mu$. For sufficiently large values of $D$, the number of particles per cluster (unit cell) is $K=1$. However, on further increasing the chemical potential,  particles bunch into clusters (also referred to as ``droplets'') which organize in a solid preserving the triangular structure. 

\begin{figure}[h]
\includegraphics[scale=0.35]{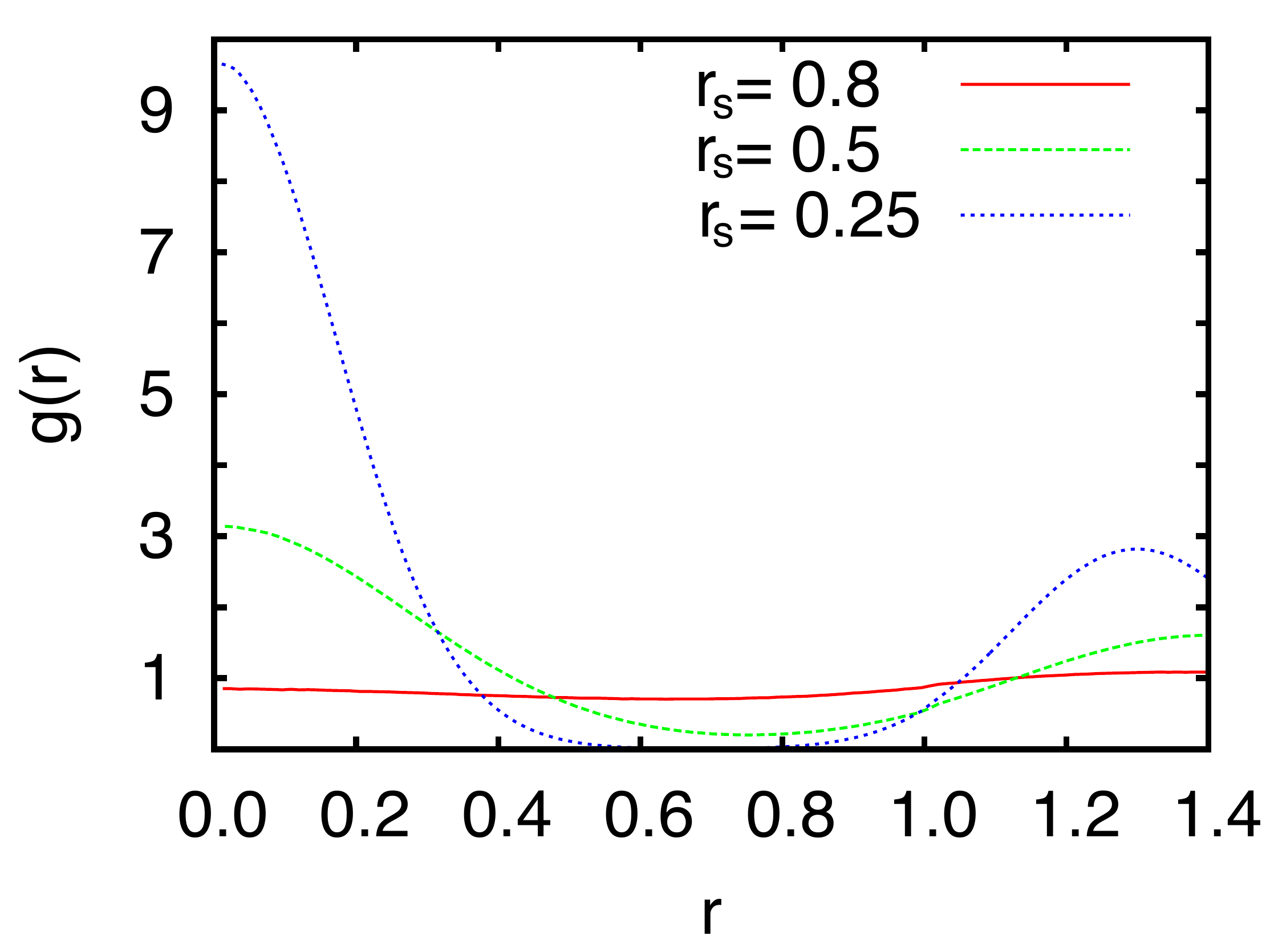}
\caption{\label{Fig_2} Ground state pair correlation function for $D=5$, at three different values of $r_s$. The solid line ($r_s = 0.8$) corresponds to a superfluid gas, the dashed one ($r_s = 0.5$) to a supersolid, and the dotted one ($r_s = 0.25$) to a non-superfluid cluster crystal.}
\end{figure}
The appearance at high density of such a cluster phase is a classical effect, directly related to the finite energy cost associated to particles being at a distance less than the soft core diameter. Indeed, it is relatively simple to estimate the number of particles per cluster at a given nominal density $r_s$, by minimizing the potential energy per particle. Cluster formation becomes favourable for $r_s <1$,  and one finds $K = \alpha/r_s^2$, where $\alpha$ is a number $\lapx 2$, {\it independent} of $D$.

For lower values of $D$, the main qualitative change is that the system crystallizes at higher densities (i.e., smaller values of $r_s$), directly into a cluster crystal with $K > 1$. 
The formation of the cluster crystal is signalled in the pair correlation function $g(r)$, as shown in Figure \ref{Fig_2}) for the case $D$ = 5. The pair correlation function is essentially featureless in the case of a gas, i.e., $g(r) \sim 1$ with only a slight depression for $r \approx 1$. For a cluster crystal, $g(r)$ develops a peak at $r = 0$, signalling multiple occupation of a single unit cell,  as well as robust oscillations, with period consistent with the lattice constant of the cluster crystal. As the density increases, clusters comprise a larger number of particles, as reflected by the increasingly greater strength of the peak at $r = 0$.  
\begin{figure}[h]
\includegraphics[scale=0.35]{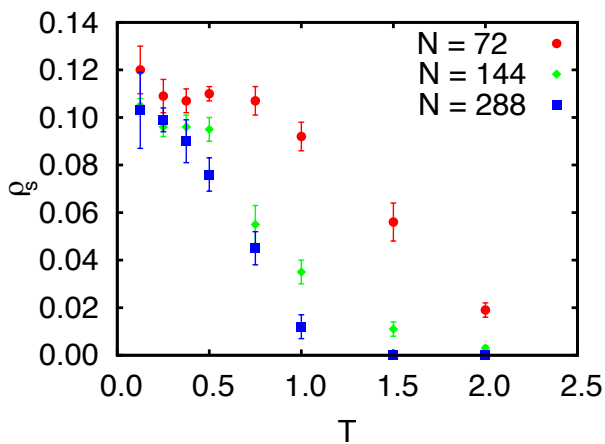}
\caption{\label{Fig_2a} Superfluid fraction of as a function of $T$ (in units of $\epsilon_\circ$) for a system of soft disks with $D$=5. Here, $\mu$ is set so that $r_s=0.5$. Data shown are for three system sizes, comprising $N$=72 (circles), $N$=144 (diamonds) and $N$=288 (squares) particles. When not shown, statistical errors are smaller than symbol sizes.}
\end{figure}

In the low $T$ limit, this system may develop phase coherence and display dissipation-less flow, besides diagonal order. We refer to this phase as  ``cluster supersolid", completely analogous to the one described in Ref. \onlinecite{cinti}. In the cluster supersolid phase, particles tunnel between adjacent clusters, and the pair correlation function takes on a finite value between successive peaks. In the insulating cluster crystal, on the other hand, $g(r)$ is essentially zero between peaks (see Figure \ref{Fig_2}). The presence of a global superfluid response, extending to the whole system,  is assessed numerically through the direct computation of the winding number.\cite{pollock} A typical result is shown in Figure \ref{Fig_2a}, for $D=5$ and $r_s=0.5$. As in any simulation study, the superfluid transition is smeared by finite-size effects, and accurate finite-size scaling analysis of the result obtained on systems comprising significantly different numbers of particles is required, in order to determine accurately the transition temperature. The results shown in Figure \ref{Fig_2a} are consistent with a superfluid transition in the Kosterlitz-Thouless universality class,\cite{kt} as expected for a two-dimensional system. 
It is worth noting that the superfluid fraction does {\it not} saturate to a value of 100\% as $T \to 0$. This is in line with the spontaneous breaking of translational invariance associated to crystalline order, as first pointed out by Leggett.\cite{leggett}

\begin{figure}

 \includegraphics[scale=0.38]{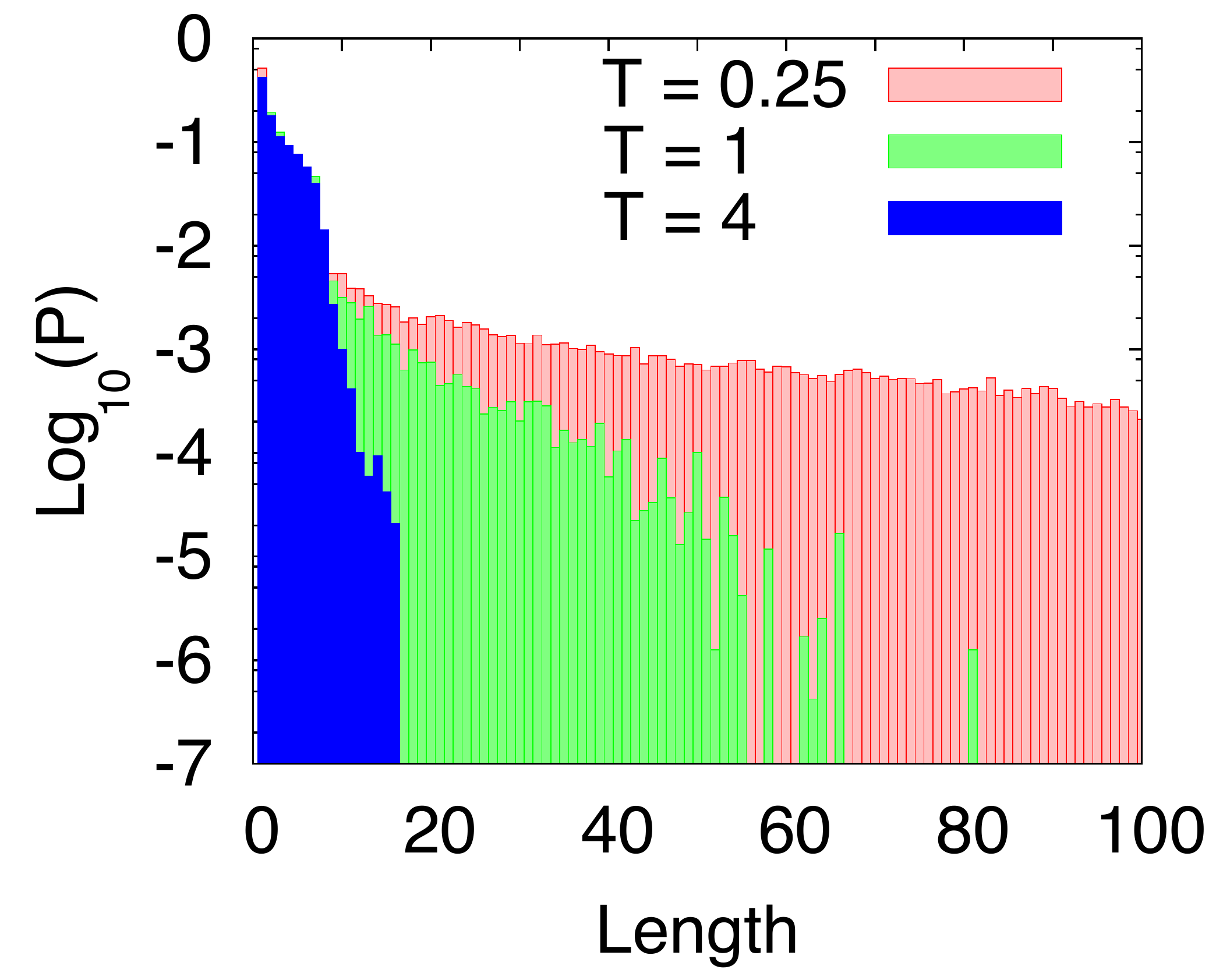}

\caption{\label{Fig_3} Frequency of occurrence $P$ of permutation of different length (i.e., involving different numbers of particles), at temperature $T$= 4 (dark boxes), $T$=1 and $T$=0.25 (light boxes). Temperatures are in units of $\epsilon_\circ$. Here, $D=5$ and $r_s=0.5$. The system is in the cluster crystal phase at all the temperatures.  }
\end{figure}

The onset of superfluidity is well known to be underlain by long cycles of exchanges of identical particles (permutations). Fig. \ref{Fig_3} shows histograms of frequency of occurrence of exchange cycles of different ``length", i.e., involving different numbers of particles, at three different temperatures for a system in the cluster crystal phase. At the highest $T$, $P$ drops sharply for cycles involving more than the number of particles per cluster (approximately 7). At the lowest $T$, exchanges involving all particles in the system set in. However, even at the highest temperature clusters are individually superfluid, even though global phase coherence does not exist.

\begin{figure}[h]

 \includegraphics[scale=0.35]{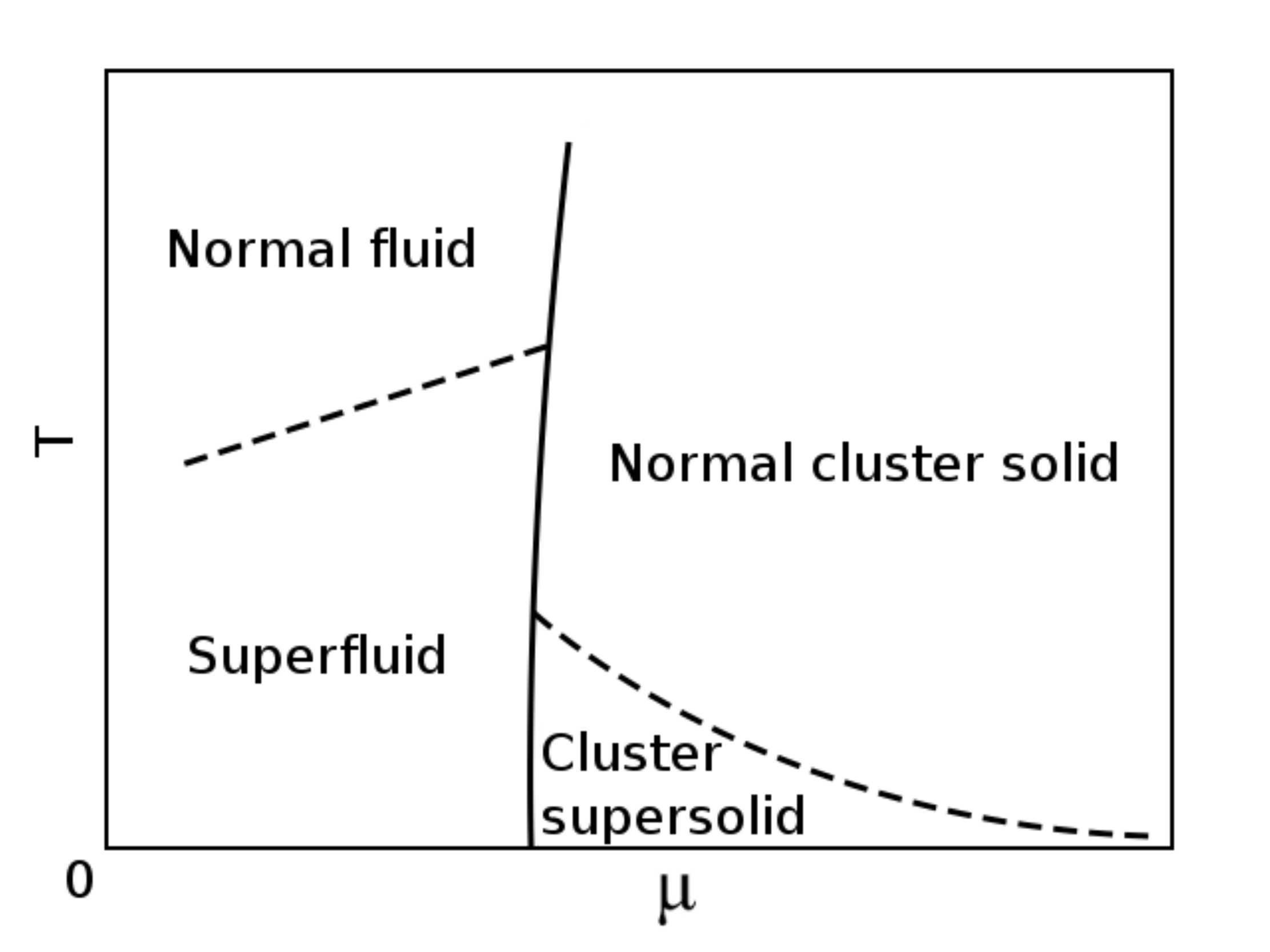}

\caption{\label{Fig_4}Qualitative low $D$ finite $T$ phase diagram of the system. Thick lines show first-order, dashed lines continuous phase transitions.}
\end{figure}

The supersolid phase described above has been observed in this work at low $T$, for $D\ \lapx 10$ and $r_s \approx 0.5$, which corresponds to a unit cell occupation $K \sim 5$. For large values of $D$ (i.e., $D \ge 10$), clusters forming a crystal are increasingly compact, i.e., particles pile up on a small spatial region, and an exponential decrease of the tunnelling rate for particles across clusters is expected, as the potential energy barrier associated with tunnelling grows linearly with $D$. The same effect takes place for $D < 10$, on increasing $\mu$, as the  number $K$ of particles per cluster becomes large, in this case the potential energy barrier for tunnelling growing linearly with $K$. Our  data at finite $T$ are consistent with an exponential decrease with $D$ or $\mu$ of the superfluid transition temperature of the cluster crystal,  but we cannot rule out continuous quantum phase transitions between a supersolid and a normal cluster crystal, driven by either $\mu$ or $D$.  The resulting schematic finite temperature phase diagram, for a value of $D$ for which a supersolid phase is observed, is shown in Figure \ref{Fig_4}. 

{\em Conclusions}. In summary, we have studied by  Monte Carlo simulations the phase diagram of  a two-dimensional system of bosons  interacting via a  repulsive, short-range soft-core potential. This system displays a low-temperature supersolid phase, wherein particles tunnel across nearest-neighbouring, multiply occupied unit cells.  The same qualitative behaviour was previously found in similar work, \cite{cinti} where, in addition to the soft core, a long-range repulsive interaction between particles was included. The results reported here show that no long-range tail is necessary for the onset of superfluidity. The most important conclusion, besides the fact that soft-core bosons seems to  be the ``minimal" model for supersolidity, is that the physics illustrated here should be observable under relatively broad conditions, if soft-core pair-wise interaction potentials can be fashioned.
%
%

This work was supported by the Natural Science and
Engineering Research Council of Canada under research grant
G121210893, and by the Italian MIUR under COFIN 07. One of us 
(MB) gratefully acknowledges hospitality of IOM-CNR at SISSA.

\end{document}